\documentclass[graybox,natbib,nosecnum]{svmult}
\bibpunct{(}{)}{;}{a}{}{,} 

\pdfoutput=1   

\usepackage{mathptmx}       
\usepackage{helvet}         
\usepackage{courier}        
\usepackage{type1cm}        

\usepackage{makeidx}         
\usepackage{graphicx}        
\usepackage{multicol}        
\usepackage[bottom]{footmisc}
\usepackage[normalem]{ulem}	
\usepackage{hyperref}  

\usepackage{soul}   

\newcommand*\aap{A\&A}

\newcommand*\aj{AJ}

\newcommand*\apj{ApJ}
\newcommand*\apjl{ApJ}

\newcommand*\araa{ARA\&A}

\newcommand*\icarus{Icarus}

\newcommand*\mnras{MNRAS}

\newcommand*\nat{Nature}

\newcommand*\planss{Planet Space Sci}

\newcommand*\psj{PSJ}

\newcommand*\ssr{Space Sci Rev}

\begin{document}

\title*{Interstellar Objects in the Solar System}
\author{David Jewitt}
\institute{David Jewitt, Dept.~Earth, Planetary and Space Sciences,  University of California at Los Angeles, 595 Charles Young Dr E, Los Angeles, CA 90095 \email{jewitt at ucla.edu}}
\maketitle

\abstract{1I/`Oumuamua and 2I/Borisov are the first  macroscopic interstellar objects to be detected in the solar system. Their discovery has triggered a tsunami of scientific interest regarding the physical properties,  dynamics and  origin of the so-called interstellar interloper population. While it is clear that a deep understanding of these issues cannot be reached from a sample of just two bodies, the emergence of this new field of astronomical study is particularly fascinating, with ramifications from planetary science to galactic dynamics. \footnote{To be published in: Handbook of Exoplanets, 2nd Edition, Hans Deeg and Juan Antonio Belmonte (Eds. in Chief), Springer International Publishing AG, part of Springer Nature.}}

\section{Introduction }
Interstellar interlopers are macroscopic bodies passing through the solar system but originating outside it.  This is a brief overview of what is known about such objects and their sources, written at the non-specialist graduate student level.  With only two known interstellar interloper examples (1I/`Oumuamua and 2I/Borisov) this is a data-limited field and, as a result, the literature is heavily loaded with speculative, model-based papers.  This review will focus on the available observations, and on a subset of the model papers that seem to be most relevant and insightful.  A more detailed review has been published in Annual Reviews of Astronomy and Astrophysics \citep{Jew23} while other reviews include \cite{Mor22}, \cite{Sel23}, and \cite{Fit23}). \\


\section{Comets Background}
Most solar system (i.e.~gravitationally bound) comets come from one of two reservoirs \citep{Don15}.  The short-period comets (SPCs, loosely defined as having orbit periods $<$200 yr) have small inclinations and modest eccentricities; they are escapees from the trans-Neptunian (``Kuiper belt'') region beyond 30 au.  They have occupied orbits beyond the known planets for the age of the solar system, but some are rendered unstable to dynamical chaos, which brings them into short-lived planet-crossing orbits ending in the inner solar system.  There are at least $\sim10^5$ trans-Neptunian bodies larger than 100 km and perhaps 10$^{10}$ to 10$^{11}$ measured down to kilometer size scales (the latter being too small and faint to be detected using existing telescopes).  The population at large perihelion distances remains especially uncertain due to observational limitations.  The current mass is $\sim$0.1 M$_{\oplus}$ (1 M$_{\oplus} = 6\times10^{24}$ kg is the mass of Earth) but the mass at formation is thought to have been several hundred times larger.

The long-period comets (LPCs have periods $>$200 yr) come from the Oort cloud.  
The orbits of LPCs are distributed isotropically, indicating that LPC comets fall from a spherical, Sun-centered cloud.  The extent of this cloud is best estimated from the distribution of the orbital semimajor axes; the cloud has a characteristic scale $\sim$50,000 au but lacks a sharp edge, with the distribution of orbital semi-major axes falling off as $a^{-2}$ \citep{Hig15}.  The population of the Oort cloud is generally reckoned to be $N_{OC} \sim10^{12}$ for radii $\ge$1 km.  However, this number is  uncertain to order of magnitude both because accurate measurements of the sizes of LPC nuclei are lacking, and because of model uncertainties. LPC nucleus densities are also unmeasured and, as a result,  the Oort cloud mass is extremely uncertain, with a nominal value $\sim$ M$_{\oplus}$ but allowable values both ten times smaller and larger.

Objects in the Oort cloud could not have formed there because of the very low densities prevailing at such large distances from the Sun.  Instead, the Oort cloud comets appear to have formed in the vicinity of the giant planets and were scattered out, perhaps as part of the later stages of planetary growth.  Without additional forces, objects scattered by the planets would either fall back into the planetary region if ejected at less than the local escape speed, or be permanently lost to interstellar space if faster.  Emplacement into bound orbits in the Oort cloud occurred because of the combined effects of gravitational perturbations from passing stars and the action of the galactic tide.  Estimates of the emplacement efficiency vary depending on the number of stars at the ejection epoch, but most fall in the range 1\% $\le f \le$ 10\% \citep{Don15}.


In fact, the sizes of LPC cometary nuclei are rarely known even to factor-of-two accuracy, leading immediately to order of magnitude uncertainty in the Oort cloud mass.  With bulk density 500 kg m$^{-3}$ \citep{Gro19}, a population of 10$^{12}$ comets each 1 km in radius would have mass $M_{OC} = 2\times10^{24}$ kg (0.3 M$_{\oplus}$) but this is surely an underestimate given that most solar system body size distributions are sufficiently flat that the mass is dominated by the largest objects. For example,  the $\sim$120 km diameter nucleus of LPC C/2014 UN271 (\cite{Ber21}; \cite{Hui22}) contains more mass than all  other observed LPCs combined.  Taking emplacement efficiency $f$ = 5\% as a rough middle value, the population of the Oort cloud indicates that the number of comets lost to the interstellar medium was $(N_{OC}/f) \sim 2\times10^{13}$, with total mass $(M_{OC}/f) \sim 4\times10^{25}$ kg ($\sim$10 M$_{\oplus}$).  Like the mass of the Oort cloud, this estimate is uncertain even to order of magnitude, but serves to indicate that substantial quantities of material were lost.

At 50,000 au, the Keplerian velocity is $V_K \sim 10^2$ m s$^{-1}$ and the orbit period is $\sim10^7$ years, which is also a measure of the free-fall time to the Sun.  As the semimajor axes increase, the effects of erosion by passing stars and the galactic tide increase, leading to the progressive depletion of objects.  For example, consider the velocity perturbation induced by a star of mass M$_{\star}$ passing the Sun and Oort cloud at distance, $d$, is

\begin{equation}
\Delta V \sim \frac{2 G M_{\star}}{d V_{\star}}
\end{equation}

\noindent in which $G = 6.67\times10^{-11}$ N kg$^{-2}$ m$^2$ is the gravitational constant and $V_{\star}$ is the relative velocity of the passing star.  Substituting representative values $M_{\star}$ = M$_{\odot} = 2\times10^{30}$ kg and $V_{\star}$ = 25 km s$^{-1}$ gives $\Delta V \sim 1$ m s$^{-1}$.  If the velocity perturbations are isotropic, then the velocity vector should be randomized by the effects of $N \sim (V_K/\Delta V)^2$ stars.  Substitution gives $N \sim 10^4$, meaning that 10$^4$ such encounters are needed to randomize the orbit.  How long does that take?  To obtain a crude but 
informative answer note that there is presently $\sim$1 star of solar mass ($\alpha$ Centauri) within 1 pc of the Sun (a larger number of smaller stars also contribute). At $V_{\star}$ = 25 km s$^{-1}$, the time for another star to cross a sphere centered on the Sun and 1 pc in radius is $t \sim 10^5$ years, meaning that the rate of similar flybys at 1pc by solar mass stars is $dN/dt \sim10^{-5}$ year$^{-1}$.  To accumulate $N =10^4$ similar flybys would then take $N/(dN/dt) \sim 10^9$ years.  This order of magnitude estimate is confirmed by numerical simulations, which show that the initially disk shaped distribution  of orbits in the Oort cloud should become randomized on timescales of a billion of years \citep{Hig15}.  This is also the approximate exponential timescale for the erosional loss of comets from the cloud \citep{Por21}.

These estimates assume that the stellar environment of the Sun has always been as it is now.  However, the Sun probably formed as part of a compact but expanding cluster containing thousands of stars \citep{Ada10}.  In this case, the external perturbation rate would be higher at early times, and the time needed to isotropize the cloud would be shorter shorter.  The proximity of stars in a birth cluster would also allow the transfer of comets from one star to another.  The Sun's Oort cloud, for example, could contain 10s of percent comets formed in the disks of other stars or even be dominated by them.  In addition, the stronger perturbations imposed by nearer cluster stars would truncate the cloud and stabilize comet orbits with much higher binding energies (smaller semimajor axes) than the 50,000 au scale of the Oort cloud \citep{Kai08}.  These bodies would define a so-called inner Oort cloud that is largely immune to direct measurement, in the sense that the LPCs are primarly derived from weakly bound orbits in the more distant regions.  Unfortunately,  uncertainties concerning the properties of the birth cluster and, especially, the relative timing of important dynamical events (ejection by planet migration, the survival and temporal evolution of the birth cluster) conspire to weaken the interpretative and predictive power of models (\cite{Kai08}; \cite{Bra13}). Improved surveys of the LPC population, in search of new evidence for surviving anistropies in the arrival directions and better measurements of the pre-approach orbits, should provide useful data from which to better reveal the properties and formation of the Oort cloud.

\section{Properties of 1I/`Oumuamua and 2I/Borisov}
\begin{figure}
\centering
\includegraphics[scale=.4]{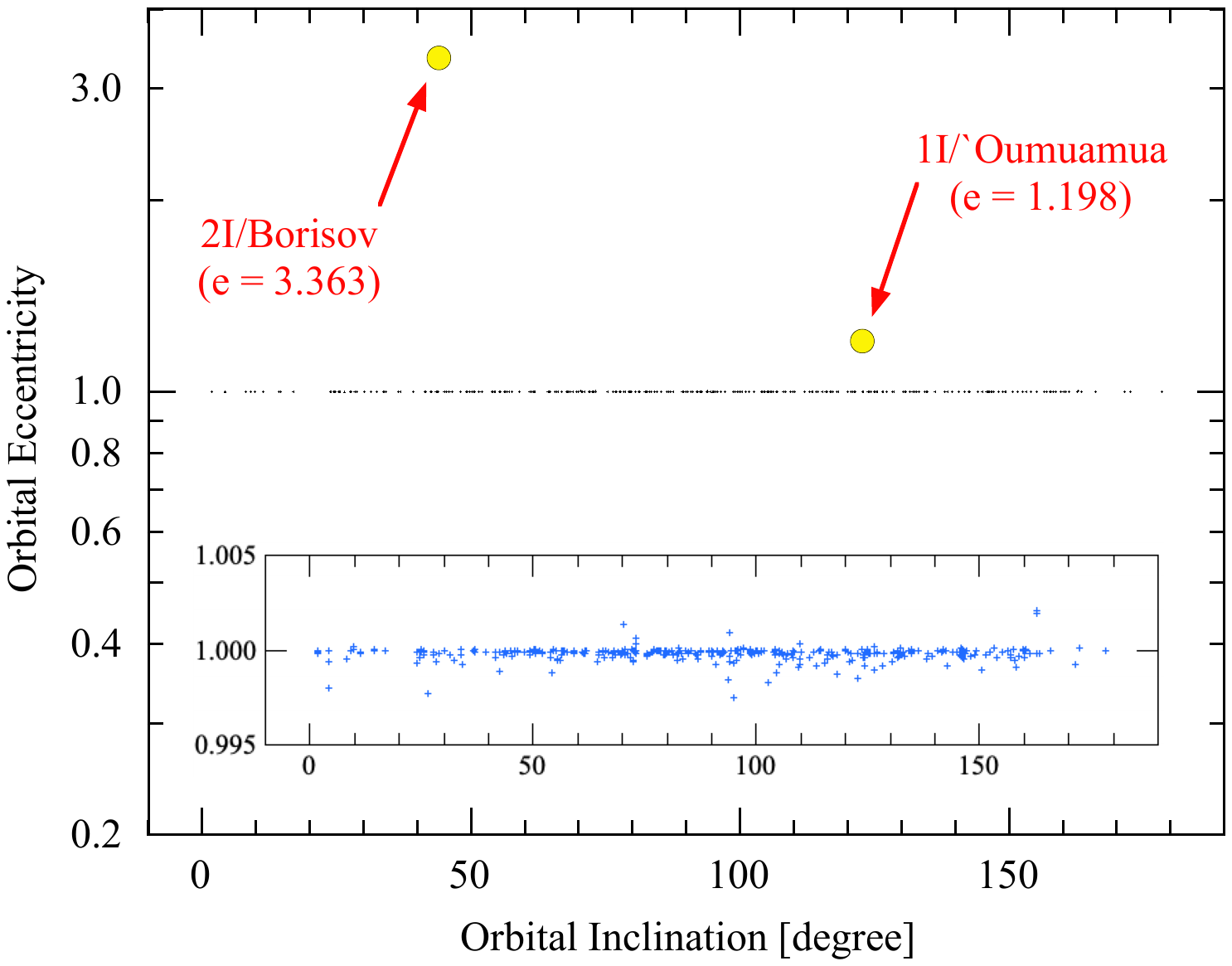}
\caption{Original orbital eccentricity vs.~inclination for long-period comets and interlopers compared.  The black band at $e$ = 1 consists of 503 comets, also shown in blue at a magnified vertical scale in the image inset.  The two interlopers stand out from all the other comets.  }
\label{LPC}       
\end{figure}

  Figure \ref{LPC} shows the  eccentricity plotted against inclination for 503 well-observed long-period comets  from  \cite{Kro20}.  The plotted elements are the so-called ``original'' values that have been corrected for the perturbing influences of the planets.  The extreme narrowness of the comet eccentricity distribution around $e$ = 1 is emphasized in blue in the inset panel.  The figure shows that  1I/`Oumuamua and 2I/Borisov differ from each other in being retrograde and prograde, respectively, but that  both  clearly stand apart from the LPCs.   The defining dynamical property of 1I/`Oumuamua and 2I/Borisov is  their orbital eccentricity (Table \ref{elements}). 

\begin{table}
\caption{Orbital Properties$^a$}
\label{elements}       
%
%
\begin{tabular}{p{2.5cm}p{1.5cm}p{1.5cm}p{1.5cm}p{1.5cm}p{2.5cm}p{2.9cm}}
\hline\noalign{\smallskip}
Object & a (au) & e & i (deg) & q (au) & $\alpha_{ng}(1)$ (m s$^{-2})$\\
\noalign{\smallskip}\svhline\noalign{\smallskip}
1I/`Oumuamua & -1.272  & 1.198  & 122.8 & 0.252 & 5.6$\times10^{-6}$\\
2I/Borisov & -0.850  & 3.363 & 44.0 & 2.009 & 1.5$\times10^{-6}$\\
\noalign{\smallskip}\hline\noalign{\smallskip}
\end{tabular}\\
$^a$ a, e, i are the orbital semimajor axis, eccentricity and inclination, respectively.  q is the perihelion distance.  $\alpha_{ng}(1)$ is the magnitude of the non-gravitational acceleration scaled to 1 au heliocentric distance.
\end{table}

\begin{figure}
\centering
\includegraphics[scale=.3]{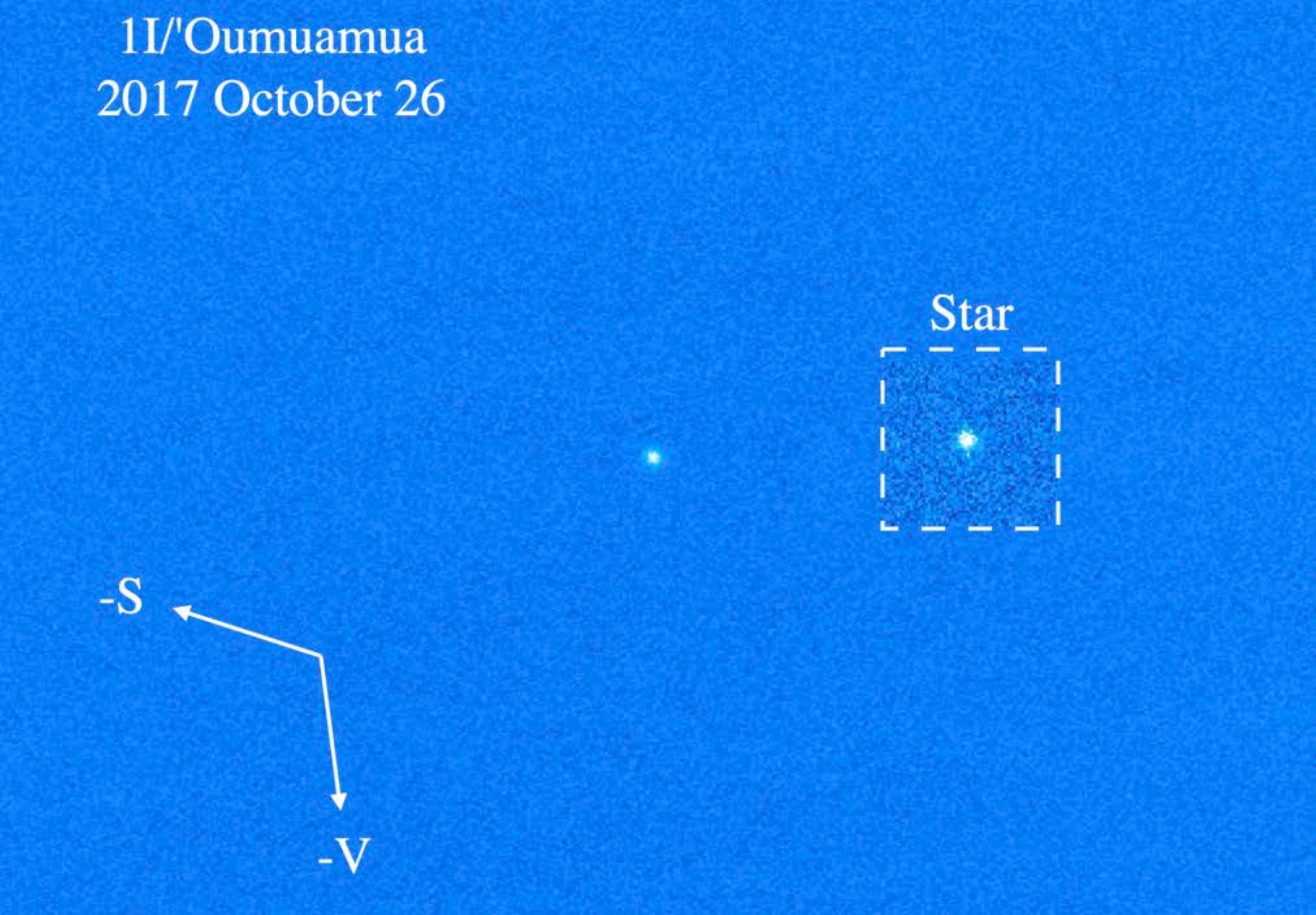}

\caption{Nordic Optical Telescope image of 1I/`Oumuamua on 2017 October 26 when at 1.40 au post-perihelion.  `Oumuamua is centered while the inset shows a field star of similar brightness for comparison.  No coma or dust trail is evident.  The width of the inset corresponds to 800 km at the 0.44 au geocentric distance to the object.  Arrows marked -V and -S show the negative projected heliocentric velocity vector and the projected antisolar direction, respectively.}
\label{Oumuamua}       
\end{figure}

\textbf{1I/`Oumuamua:}  The distinctive physical properties of 1I/`Oumuamua include an extreme rotational lightcurve range of $\sim$2.5 magnitudes, a rotation period near 8 hours,  the possibility of small differences between successive rotations of the body that are suggestive of excited (non principal axis) rotation, and the complete absence of evidence for ejected material in the form of either a dust or a gas coma (\cite{Jew17}; \cite{Mee17}; \cite{Dra18}, c.f. Figure \ref{Oumuamua}).  The absence of coma makes direct study of the nucleus possible. 
 Unfortunately, interpretations of the body shape from the lightcurve are non-unique.  For example, the possibility that the large range is caused by azimuthal albedo variations cannot be formally rejected, other than by arguing that examples of such large albedo variability in the solar system are vanishingly rare.  Assuming albedo uniformity, the lightcurve was at first interpreted as caused by a prolate, rocket-like shape with an axis ratio of 10:1 (later reduced to 6:1 following allowance for scattering effects in the non-zero phase angle observations \cite{Dra18}).  This extreme elongation is distinct from the main-belt asteroids, where the average axis ratios of the best-fit tri-axial ellipsoids are nearer 1:1.4:2.  While a highly elongated body shape cannot be rejected, it is statistically unlikely for such a body to have its spin axis oriented such that observers on Earth would see the full lightcurve.  For this reason, the shape is widely accepted to be oblate or disk-like, like a hamburger with an axis ratio $\sim$6:6:1 and rotating far from the minimum energy (i.e.~maximum moment of inertia) configuration \citep{Mas19}. Slight non-repeating features in the lightcurve, if not due to measurement errors or weak, unresolved cometary activity, could indicate non-principal axis rotation \citep{Dra18} and have been modeled as such \citep{Tay23}.

Optical photometry measures the product of the albedo with the scattering cross-section of the nucleus, but does not provide either quantity separately.  The albedo of 1I/`Oumuamua remains unmeasured and so the cross-section (and size) cannot be uniquely determined.  Estimates of the circle-equivalent radius of 1I/`Oumuamua span the range from $\sim$50 m to $\sim$110 m.  For the following, a nominal geometric albedo of 0.1 is assumed, giving a scattering cross-section at mid-light  corresponding to a circle  $\sim$ 80 m in radius.  This value, which is itself uncertain by a factor $\sim$2, is an order of magnitude smaller than the kilometer scale, well-studied  solar system comets.  Taking account of the shape, 1I/`Oumuamua can be considered as a disk-like body about 110$\times$110$\times$18 m. 

Spectroscopic observations revealed none of the common cometary gases (CN, C$_2$, C$_3$, OH, CO, CO$_2$) in `Oumuamua, with empirical upper limits to the production rates $\ll$1 kg s$^{-1}$ other than for water ($\ll$50 kg s$^{-1}$).  Limits to the rate of dust production are more model dependent but also very small.  Mass loss rates in micron sized particles  exceeding 10$^{-4}$ kg s$^{-1}$ \citep{Jew17} to 10$^{-3}$ kg s$^{-1}$ \citep{Mee17} would have produced detectable dust coma, where none was seen. \\

\begin{figure}
\centering
\includegraphics[scale=.3]{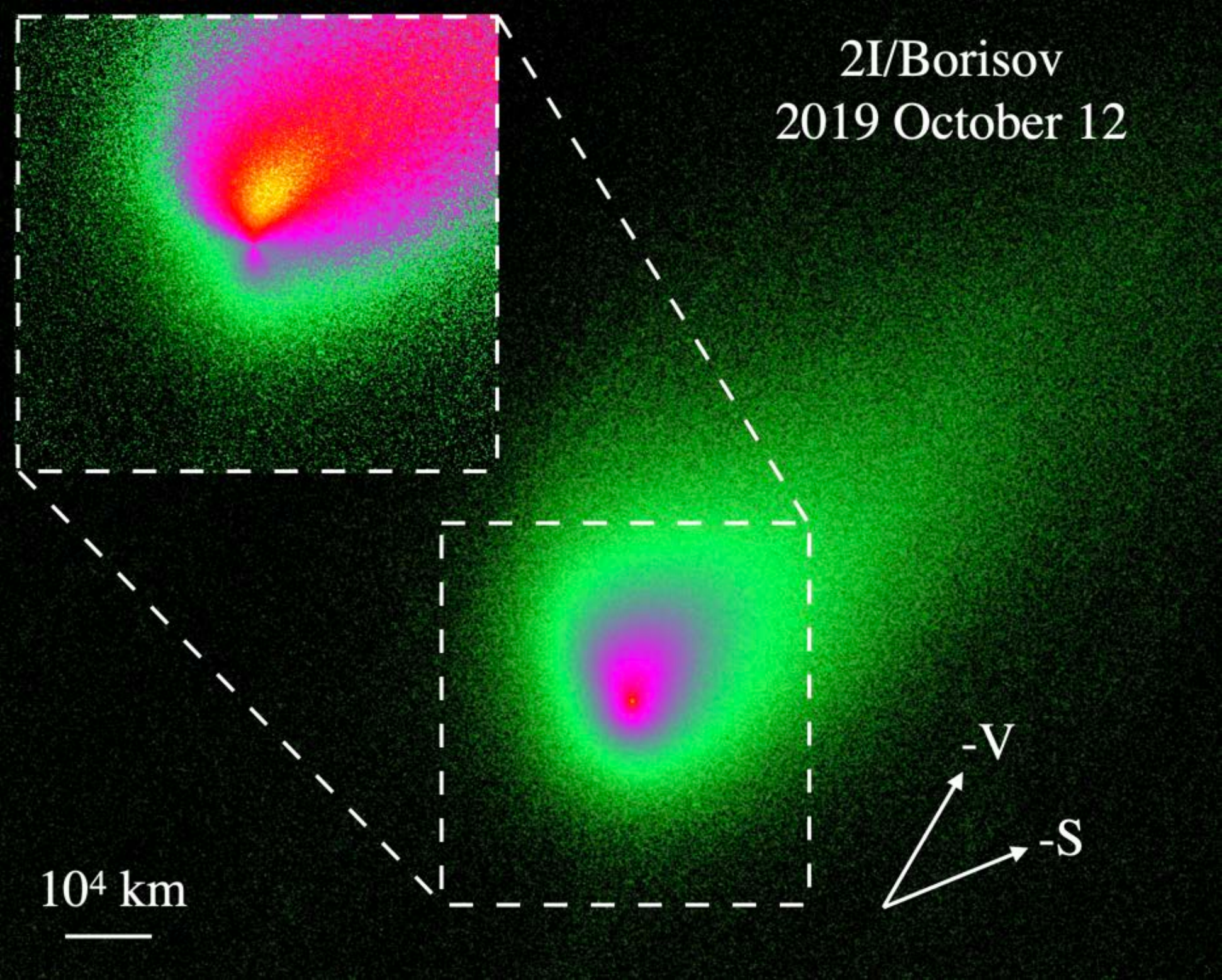}

\caption{Hubble Space Telescope image of 2I/Borisov on 2019 October 12 when at 2.37 au pre-perihelion.  The image at lower right shows the broadband optical continuum, contributed by ejected dust particles.  The inset at upper left shows the same image divided by a coma model in order to emphasize azimuthal structures  caused by anisotropic outgassing (``jets'') and radiation pressure.  Arrows marked -V and -S show the negative projected heliocentric velocity vector and the projected antisolar direction, respectively.}
\label{Borisov}       
\end{figure}

\textbf{2I/Borisov:}  
 Unlike 1I/`Oumuamua, 2I/Borisov showed prominent, comet-like activity in both pre- and post-perihelion observations (c.f.~Figure \ref{Borisov}; \cite{Jew19}; \cite{Guz20}). The coma of 2I/Borisov was sufficiently bright that the nucleus could not be directly studied.  
 Non-detection even at the 0.08" resolution of Hubble Space Telescope set an upper limit to the radius $r_n <$ 500 m, while an argument based on Equation \ref{dmbdt} sets a lower limit $r_n >$ 200 m \citep{Jew20b}.  Because the nucleus was not detected, there are no useful constraints on its rotation period or shape.

 Models of the effect of solar radiation pressure on the morphology show that the coma was dominated by large particles, with effective  sizes $\ge$100 $\mu$m,  ejected slowly ($\le$9 m s$^{-1}$) and with a dust mass loss rate $\sim$35 kg s$^{-1}$ \citep{Kim20}.  Several gas species were detected, including H$_2$O \citep{Xin20} and CO \citep{Bod20}, with production rates of several $\times$10 of kg s$^{-1}$ at the 2 au perihelion.  The ratio of CO to water production rates, $Q_{CO}/Q_{H_2O} \sim$ 1 (\cite{Bod20}; \cite{Cor20}) was comparable to the largest values measured in solar system comets near the same heliocentric distance and far larger than the average value ($\sim$0.04).  The high abundance of this very volatile ice is consistent with the formation  of the nucleus of 2I/Borisov at very low temperatures (probably $\le$30 K) and with the even lower average temperatures (10 K or less) experienced while in interstellar space.

\section{Non-Gravitational Motion}
The motions of both 1I/`Oumuamua and 2I/Borisov show the existence of a non-gravitational acceleration (NGA: c.f.~Table \ref{elements}), (\cite{Mic18}; \cite{Hui20}) for which several explanations have been proposed.

\textbf{Recoil from Outgassing:}
 In solar system comets, NGA is common and results from the anistropic ejection of mass, usually caused by the preferential sublimation of ice from the hot dayside of the cometary nucleus.  As a result, the principal component of the NGA acts in the antisolar direction.  Within the uncertainties of measurement, outgassing rates from 2I/Borisov are  sufficient to explain its measured NGA.  However, no outgassing was detected from 1I/`Oumuamua, raising a puzzle concerning the cause of its acceleration.

By force balance, the mass loss rate from a spherical nucleus of radius, $r_n$, needed to induce a non-gravitational acceleration, $\alpha_{ng}$, is

\begin{equation}
\frac{dM}{dt} = \left(\frac{4\pi\rho_n r_n^3}{3 k_R V_s}\right)\alpha_{ng}
\label{dmbdt}
\end{equation}

\noindent where $\rho_n$ is the density, $k_R \sim0.5$ is a coupling constant to represent the fraction of the outflow momentum delivered to the nucleus (isotropic mass loss would have $k_R$ = 0, perfectly collimated mass loss would have $k_R$ = 1), and $V_g$ is the speed of the sublimated ice molecules.  Adopted values include a comet-like $\rho_n$ = 500 kg m$^{-3}$ \citep{Gro19} and a gas outflow speed $V_g$ = 500 m s$^{-1}$, also based on measurements of comets.  

The measured acceleration of 2I/Borisov scaled to the perihelion distance $r_H$ = 2 au by the inverse square law is $\alpha_{ng} = 3.8\times10^{-7}$ m s$^{-2}$.  With nucleus radius 200 m to 400 m and other parameters as listed above, this acceleration corresponds to a mass loss rate $dM/dt \sim$ (25 to 200) kg s$^{-1}$, broadly consistent with measured gas production rates \citep{Bod20}.  

The result is very different for 1I/`Oumuamua.  The measured NGA at $r_H$ = 1 au is $\alpha_{ng} = 6\times10^{-6}$ m s$^{-2}$ which, from Equation \ref{dmbdt}, gives $dM/dt$ = 25 kg s$^{-1}$. No evidence for mass loss at this level was detected.    One possibility is that the mass loss could be in the form of a spectroscopically undetected gas, and several possible molecules have been proposed.

The dominant volatile in solar system comets is water. The empirical limit to production in 1I/`Oumuamua is $dM/dt <$ 30 kg s$^{-1}$, but the volatility of water ice is low enough that a body of Oumuamua's size cannot generate even 1/10th of this amount. Water cannot accelerate 1I/`Oumuamua unless the nucleus density is $\ll$500 kg m$^{-3}$ \citep{Sek19}.

\cite{Sel20} suggested that 1I/`Oumuamua might consist of solid hydrogen. While H$_2$ is cosmically abundant, it is also extremely volatile and would be difficult to accrete at even the lowest temperatures found in molecular cloud cores.  Even if this were possible, the thermal stability of solid H$_2$ during transit across the open interstellar medium is doubtful.  

\cite{Ber23}  circumvented the thermal stability problem  by assuming that the H$_2$ is trapped within relatively refractory amorphous water ice formed radiolytically by prolonged cosmic ray bombardment of the nucleus.  An unaddressed problem is that H$_2$ release begins  at  very low temperatures ($\sim$15 K), which are to be found meters beneath the physical surface of an incoming `Oumuamua-like body, according to these authors.  Temperatures at depth are effectively decoupled from the surface temperature because the thermal diffusion time is very long compared to the diurnal timescale.  For example, at depth $\ell$ = 1 m  
 and with comet-like diffusivity $\kappa$ = 10$^{-8}$ m$^2$ s$^{-1}$, the diffusion timescale is $\tau \sim \ell^2/\kappa \sim 10^8$ s, or about 3 years, whereas the diurnal timescale is three orders of magnitude shorter at only $\tau_d \sim$ 8 hours. Under these circumstances, H$_2$ escape would not be restricted to the sunward hemisphere and the net recoil force would be smaller and not be antisolar.

\cite{Des21} proposed that solid nitrogen (as N$_2$ ice) is the driver of 1I/`Oumuamua's non-gravitational motion. N$_2$ is spectroscopically inert and so would naturally escape detection.  The main question is how would an N$_2$ iceberg form?  Noting that Pluto's surface is partly coated in solid N$_2$, these authors suggested that impacts into Pluto analogs in other systems might produce N$_2$ rich fragments. The galactic abundance of Pluto-like N$_2$ ice lakes, their rates of impact excavation and the cosmic efficacy of this mechanism are all unclear. 

Carbon monoxide (CO) is abundant in solar system comets (at levels $\sim$1\% to 10\% that of water) and might be expected to be more abundant in frigid interstellar bodies like 1I/`Oumuamua.  Unfortunately, the observational limit to CO outgassing ($dM/dt < 4\times10^{-2}$ kg s$^{-1}$; \cite{Sel21}) is far below the value required to account for the non-gravitational acceleration.

An additional problem relevant to all outgassing models is that of the non-detection of dust in `Oumuamua.  Cometary dust dominates the scattering cross section relative to gas (visible through resonance fluorescence) even when the dust and gas production rates are comparable.  Observational limits to  mass loss in micron-sized dust particles in 1I/`Oumuamua are tiny (e.g.~$dM/dt < 10^{-4}$ kg s$^{-1}$ \citep{Jew17} to $<10^{-3}$ kg s$^{-1}$ \citep{Mee17}), and orders of magnitude below the rate needed to explain the NGA.  The cross-section per unit mass of particles of radius $a$ varies $\propto 1/a$ such that dust could be ``hidden'' if the mean particle size were unusually large (\cite{Mic18}).  However, large particles are difficult to accelerate in a weak gas flow, and reliable models of this process do not exist.  Some near-Earth objects (notably, the $\sim$300 m diameter 523599 (2003 RM)) show NGA corresponding to mass loss at a few $\times10^{-3}$ kg s$^{-1}$ without detectable coma or outgassing \citep{Far23}.

\textbf{Radiation Pressure:}
Another possibility is that the non-gravitational acceleration of `Oumuamua results not from anisotropic outgassing, but from the action of solar radiation pressure \citep{Bia18}.  The flux of sunlight at heliocentric distance $r_H$ is $F_{\odot} = L_{\odot}/(4\pi r_H^2)$, where $L_{\odot} = 4\times10^{26}$ W is the solar luminosity.  The corresponding radiation pressure is $F_{\odot}/c$ where $c = 3\times10^8$ m s$^{-1}$ is the speed of light.  Applied to a perfectly absorbing body of cross section, $\Sigma$, and mass, $M$, the resulting acceleration is

\begin{equation}
\alpha_{ng} = \frac{L_{\odot}}{4 \pi r_H^2 c} \frac{\Sigma}{M}.
\label{alpha}
\end{equation}

\noindent The non-gravitational acceleration of `Oumuamua scaled to $r_H$ = 1 au, is $\alpha_{ng} \sim 6\times10^{-6}$ m s$^{-2}$ giving, by Equation \ref{alpha}, $\Sigma/M \sim$ 1 m$^2$ kg$^{-1}$.

Two geometries have been proposed to account for this low value of $\Sigma/M$.   
For a sphere of density $\rho$, and radius $a$, $\Sigma/M = 3/(4 \rho a)$.  Then, substituting radius $a$ = 80 m, the implied bulk density is $\rho \sim 10^{-2}$ kg m$^{-3}$, or about 10$^5$ times less dense than water.  This incredibly low density, if real, would imply a highly porous, possibly fractal structure (\cite{Mor19}, \cite{Luu20}).  1I/`Oumuamua is not accurately represented by a sphere, but comparably low densities are implied for any ellipsoidal body-shape consistent with the optical lightcurve.

\begin{figure}
\centering
\includegraphics[scale=.4]{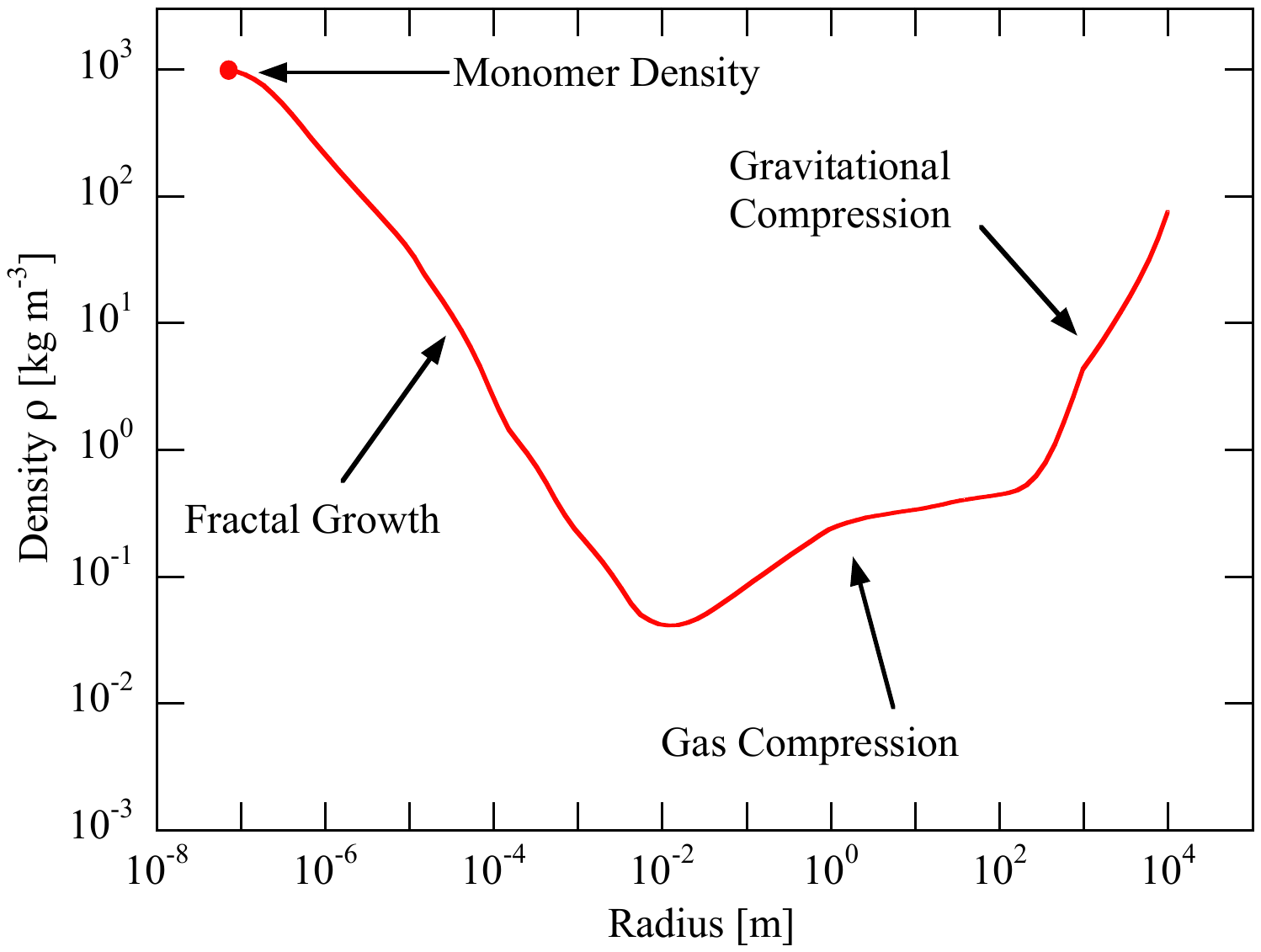}
\caption{Schematic showing bulk density as a function of object size for structures consisting of aggregated monomers, based on the discussion by \cite{Kat13}.  The monomer density and size are $\rho$ = 10$^3$ kg m$^{-3}$ and 0.1 $\mu$m, respectively.  Fractal structure produces lower and lower densities as the size increases to centimeter scale.  Thereafter the aggregates are compressed towards higher densities at first by gas pressure and then by gravitational self-compression.  }
\label{Kataoka}       
\end{figure}

 For comparison, the lowest density aerogels have $\rho$ = 0.2 kg m$^{-3}$ \citep{Mec12}.  In this context, densities of solids as low as 10$^{-2}$ kg m$^{-3}$ (10$^2$ times less dense than air) seem implausible.  However, the formation of aggregated particles might naturally lead to an open, fractal structure, with a bulk density that decreases as the size of the aggregate increases.  \cite{Kat13} present intriguing results concerning the growth and densification of fractal aggregates under static loads.  They find that minimum densities, $\rho \sim 10^{-1.5}$ kg m$^{-3}$, are reached at centimeter scales, after which aggregates are compressed first by the pressure of the gas in which they are embedded and later, at sizes $\ge$ 10$^2$ m, self-compressed by their own gravity. The resulting dependence of density on particle size is shown in Figure \ref{Kataoka}.  

The applicability of this plot in the context of the interstellar interlopers is questionable; the density curve assumes very gentle particle accretion and neglects compression by planetesimal impact in the protoplanetary disk, as well as other physical processes.  The Figure does provide a useful reminder, however, that materials may form and survive in space with properties unlike those of materials naturally occurring on Earth; the ultra-low densities inferred for `Oumuamua should not be dismissed out of hand.

Radiation pressure could also accelerate `Oumuamua without reliance on a very low density if it is in the form of a disk, for which the cross-section/mass ratio is $\Sigma/M = 1/(\rho d)$, where $d$ is the thickness. Then $\Sigma/M \sim$ 1 m$^2$ kg$^{-1}$ coupled with a comet-like density ($\rho$ = 500 kg m$^{-3}$, \cite{Gro19}), yields $d \sim 10^{-3}$ m.  In this interpretation, `Oumuamua would be a millimeter thick sheet with an effective cross-section equal to that of an 80 m radius circle.  Similarly thin, large structures are unknown in nature but are reminiscent of the aluminized plastic (``Mylar'') sheets used in solar sails \citep{Bia18}.

\textbf{Rotational Spin-Up:}
A final issue for 1I/`Oumuamua concerns the effect of outgassing on the spin of the body.  Outgassing exerts a torque, leading to a change in the spin which is ultimately limited by rotational breakup, when centripetal forces exceed gravitational and cohesive forces holding the body together.  The timescale for changing the spin (ignoring changes in the spin direction) is

\begin{equation}
\tau = \left(\frac{16\pi^2}{15}\right) \left(\frac{\rho a^4}{k_T V_s P}\right) \left(\frac{1}{\overline{\dot{M}}}\right)
\label{tau1}
\end{equation}

\noindent in which $V_s$ (m s$^{-1}$) is the speed of the escaping gas molecules, $P$ (s) is the initial rotation period and $\overline{\dot{M}}$ (kg s$^{-1}$) is the average mass loss rate \citep{Jew21}.  The timescale is a strong function of body radius, giving short spin-times for small outgassing bodies.  Measurements of comets show  that a fraction $k_T$ = 0.7\% of the outflow momentum goes into changing the spin \citep{Jew21}.  Observations of well-characterized cometary nuclei, give the empirical relation

\begin{equation}
\tau \sim 100 a^2
\label{tau}
\end{equation}

\noindent with $\tau$ expressed in years, $a$ in km, and the relation strictly applies to comets with perihelia near Earth's orbital distance.  Substitution of $a$ = 0.08 km for `Oumuamua into Equation \ref{tau} gives $\tau \sim$ 0.6 yr.  If the outgassing rate from 'Oumuamua is large enough to explain the non-gravitational acceleration, $\overline{\dot{M}}$ = 25 kg s$^{-1}$, then Equation \ref{tau1} gives $\tau \sim$ 1 month.  These short timescales are comparable to, or less than, the fly-through time in the terrestrial planet region, which suggests that `Oumuamua's rotation could be substantially affected by outgassing torques.  Indeed, small deviations from strict periodicity in the lightcurve have been interpreted as evidence for excited rotation (\cite{Dra18}; \cite{Tay23}).  

\cite{Raf18} found an even shorter timescale for `Oumuamua and suggested that its nucleus should have disrupted in the absence of material strength.  Indeed, it is conceivable that `Oumuamua is the surviving fragment of an object that disintegrated before the first observations.  However, calculations indicate that the cohesive strengths of large fractal aggregates are easily large enough to withstand rotational stresses produced by `Oumuamua's current $\sim$8 hour rotation \citep{Fle19}. Within the numerous uncertainties regarding the physical properties of this object (notably the tensile strength), survival against rotational disruption cannot be ruled out.

For 2I/Borisov, Equation \ref{tau} with $200 \le a \le$ 500 m  gives $4 \le \tau \le$ 25 yr. The longer timescale for Borisov argues against strong spin-up during its planetary flyby. A fragment  observed on 2020 March 30 (at $r_H$ = 3.2 au, outbound)  could have been rotationally ejected \citep{Jew20a} but could also have been expelled by the build up of pressure exerted by sublimating near-surface volatiles.

\section{Interloper Population and Ages}
The interloper population is practically collisionless.  If we represent the galaxy as a disk of radius $R_g$ containing $N_{\star}$ stars, the probability of an interloper-star collision upon traversing the galactic mid-plane is given by the effective optical depth, $\theta \sim N_{\star} (R_{\star}/R_g)^2$, where $R_{\star}$ is the average star radius.  Setting $N_{\star} \sim 10^{11}$, $R_g = 10$ kpc and $R_{\star} = 10^9$ m, gives $\theta \sim 10^{-12}$.  A given interloper could pass through the galactic plane 10$^{12}$ times without a collision and so would have a negligible chance of impacting a star within the age of the universe.  Along the same lines, approaches to within 5 AU of a star (the approximate distance from a Sun-like star at which water ice sublimates) correspond to $\theta \sim 10^{-6}$.  The flybys of 1I/`Oumuamua and 2I/Borisov are almost certainly  the first stellar approaches made by either object since their ejection into interstellar space.  It is therefore reasonable to consider 1I and 2I as thermally unprocessed bodies when first approaching the Sun.

\textbf{Population:} 1I/`Oumuamua was discovered as part of a relatively well characterized sky survey, allowing a useful estimate of the interloper population.  Both order of magnitude (\cite{Jew17}; \cite{Mee17}) and more detailed estimates \citep{Do18} give the number density of `Oumuamua-like bodies $N_1 \sim$ 0.1 au$^{-3}$ ($\sim$10$^{15}$ pc$^{-3}$).  Representing the planetary region of the solar system as a sphere of radius 30 au (corresponding to Neptune's orbit) gives an enclosed volume $V \sim10^5$ au$^3$.  Then, $N = N_1 V \sim10^4$  `Oumuamua-like bodies must lie within the planetary region at any instant \citep{Jew17}.  Furthermore, `Oumuamua crossed the planetary region in about 10 years from which a flux of similar bodies $dN/dt \sim10^3$ year$^{-1}$, or 3 day$^{-1}$, may be inferred.

The galactic population can be crudely estimated by representing the Milky Way as a disk 10 kpc in radius and 1 kpc thick (i.e.~volume $10^{27}$ au$^3$).  If $N_1 \sim$ 0.1 au$^{-3}$ applies uniformly across this disk, then the number of `Oumuamua-like bodies in the galaxy must be  $\sim10^{26}$, with a combined mass $\sim$ 10$^{10}$M$_{\oplus}$, corresponding to $\sim$0.1 M$_{\oplus}$ per star. This is small compared to the $\sim$10 M$_{\oplus}$ per star estimated to be lost by the Sun by scaling from the Oort cloud population and emplacement efficiency.

On the other hand, the interloper mass is likely to be a strong underestimate of the true value because the size distributions of macroscopic planetary bodies (asteroids, comets) are typically  ``top heavy'', with most mass in the largest bodies.  For example, the largest asteroid (1 Ceres) contains about 40\% of the mass in the asteroid belt.  The largest detected Oort cloud comet (120 km diameter C/2014 UN271) contains more mass
than all the other known Oort cloud comets combined. 

2I/Borisov was discovered in an informal survey of the sky the parameters of which (depth, area covered, cadence) have not been published.  As a result, it is not possible to extrapolate from this remarkable discovery to estimate the population of 2I-sized bodies.  The problem is actually worse because 2I/Borisov was active and resolved, and quantification of detection sensitivity for diffuse objects is notoriously difficult even when the point-source sensitivity is known.  

\textbf{Ages:}
Interloper lifetimes in the interstellar environment are not limited by near approaches to, or collisions with, stars.  However, the random velocities of interlopers are excited by gravitational scattering from giant molecular clouds and from density structures associated with the spiral arms of the galaxy. The result is a progressive increase in their velocity dispersion, $\Delta V$, with time, $t$, roughly according to $\Delta V \propto t^{1/2}$ (as befits a random scattering process).  Independent of any detailed model, the relative speeds of `Oumuamua (26 km s$^{-1}$) and Borisov (32 km s$^{-1}$) thus suggest that the former is dynamically the younger of the two.  Very approximate dynamical ages are $\sim10^8$ yr and $\sim10^9$ yr, for these two bodies \citep{Hal20}.  
If ejected from another planetary system with an excess velocity of 1 km s$^{-1}$, a 10$^8$ year travel time would correspond to $\sim$100 pc, meaning that the source system of `Oumuamua might be comparatively close.  This realization has prompted numerous attempts to locate the origin of this body.  A definitive location cannot be specified, but the  Carina and Columba moving stellar groups are suitably located and of similar (45$\pm$10 Myr) age (\cite{Gai18}, \cite{Hal20}). It is possible that a significant fraction of the interlopers are contributed by focused streams from such nearby sources, instead of being part of a more uniform galactic flux, in which case the galaxy-wide population estimate mentioned above may be unreliable.  The greater age and larger distance traveled by 2I/Borisov render hopeless any attempt to identify its site of origin.

\section{Origins}
Several possible origins have been suggested for the population of macroscopic interstellar bodies.  In principle, Oort cloud comets can acquire hyperbolic trajectories by being scattered by passing low mass stars and free-floating planets. However, models show that this process offers an implausible explanation for the highly eccentric orbits of 1I/`Oumuamua and 2I/Borisov \citep{Hig20}.   The likely origins, discussed below, all involve formation far beyond the Sun's region of gravitational control. 

\textbf{Protoplanetary Disk Ejection:}
As the planets grow they exert an ever larger influence over the future dynamical evolution of the planetesimal disk.  Planetesimals interacting gravitationally with the planets, if they do not collide and contribute to planetary growth, can be sling-shot accelerated out of the solar system.  The maximum possible speed increase in a sling-shot interaction is of the same order as the planetary escape speed, $V_e$.  Therefore, the efficacy of this process depends on the relative magnitudes of $V_e$ and the escape speed from the planetary system, $V_{SS}$, measured at the orbital distance of the planet.  The relevant  metric is  the Safronov number, $S = V_e^2/(2V_{SS}^2)$, with escape occuring for $S >1$ but not otherwise, at least in single-scattering events.   The Safronov number may be written

\begin{equation}
S = \left(\frac{M_P}{M_{\star}}\right)  \left(\frac{a_P}{R_P}\right)
\label{Saf}
\end{equation}

\noindent where $M_P$ and $R_P$ are the mass and radius of the planet, $a_P$ is the orbital semimajor axis, and $M_{\star}$ is the mass of the star.   

\begin{figure}
\centering
\includegraphics[scale=.4]{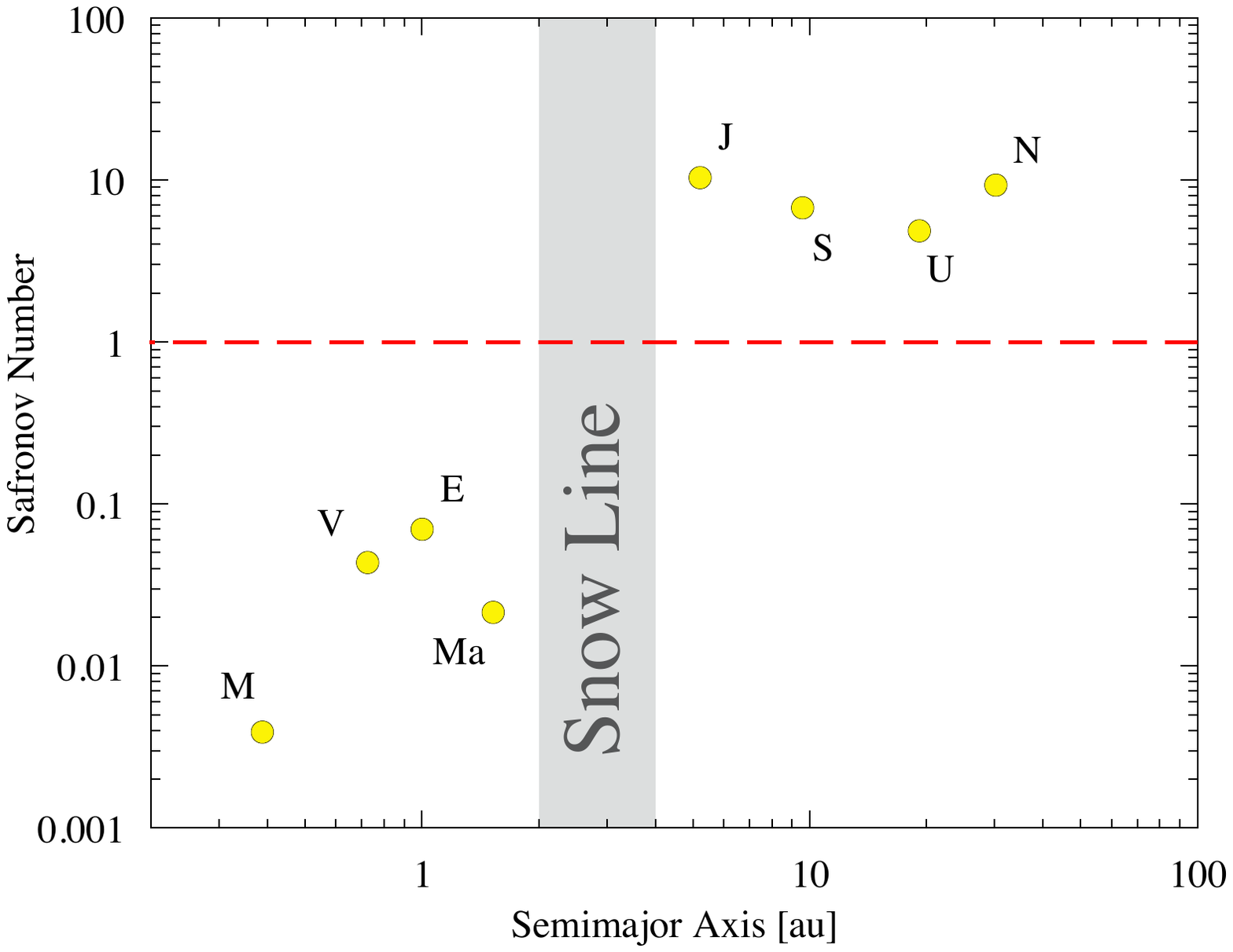}
\caption{Safronov number, $S$, for solar system planets.  The dashed red line marks $S$ = 1, above which scattering ejection to escape is posible.  The approximate location of the snow line is marked by a vertical grey band. The four giant planets can eject to  interstellar space.  The four terrestrials cannot.}
\label{safronov}       
\end{figure}

Figure \ref{safronov} shows $S$ for the planets of the solar system, emphasizing the role of planetary mass and orbit size in the ejection process.  The four giant planets have $S > 1$ and are  capable of scattering bodies to the interstellar medium while the four terrestrial planets are not.  As a result, the vast majority of ejectable bodies lie beyond the snow line (the distance beyond which water molecules freeze into solids).  This is why the Oort cloud is dominated by ice-rich comets not by refractory asteroids, although a small rocky component might exist \citep{Wei97}.  Note that because the density of the protoplanetary disk decreases outwards, Jupiter dominates the ejection of material from the solar system \citep{Hah99}.  Large Safronov number planets are preferentially found in the outer regions of known exoplanet systems (Figure 15 of \cite{Jew23}), further suggesting that the interloper population should be dominated by volatile rich comet-like, not asteroid-like,  bodies.

Dissipation of the protoplanetary disk of a star takes perhaps $\sim$5 Myr, while planetary migration may continue to excite the orbits and stability of remaining planetesimals for perhaps $\sim10^8$ years \citep{Mor18}.  Long after these processes are complete, external perturbations and stellar evolution play roles in the continuing loss to the interstellar medium of initially bound objects.  In the case of the solar system, erosion by passing stars leads to the loss of 25\% to 65\% of the Oort cloud mass on the $\sim$10 Gyr main-sequence lifetime \citep{Han18}.    

\textbf{Post-Main Sequence Ejection:}
Stars more massive than $\sim$0.9M$_{\odot}$ exhaust their core hydrogen within the age of the universe and then evolve away from the main sequence.  Post main sequence evolution is comparatively rapid, and typically involves the loss of stellar mass  through a combination of both steady  and  pulsed stellar winds.  The orbits of planetary and other bodies around these stars can be destabilized by this loss of mass.  Very distant orbits of Oort cloud bodies can simply detach from the host star but more tightly bound orbits can also be destabilized by being cast into planet and even star-crossing orbits, leading to ejection \citep{Lev23}.  Such bodies, called ``Jurads'' (after the late  Michael Jura, \cite{Han17}) might be compositionally distinct from ice-rich planetesimals released from a protoplanetary disk as a result of thermal processing induced by long term exposure to the parent star.  While very uncertain, calculations suggest that Jurads should be outnumbered by interstellar comets ejected from the protoplanetary disks of young stars, but still might be numerous enough to be detected by planned telescopic surveys \citep{Lev23}.

\textbf{Tidal Shredding:}
Depending on their density and strength, bodies passing closer to a star or planet than the Roche radius can be tidally shredded.  Because a single object can be shredded into many smaller bodies, this counts as a potential mechanism for the production of interlopers, provided the resulting fragments escape from the system.  The best known example of tidal shredding in the solar system is that of comet P/Shoemaker-Levy 9, which was sheared into $\sim$20 fragments during a close (1.6 planet radii) approach to Jupiter (\cite{Wea95}, \cite{Asp96}). However, no fragments escaped and all impacted Jupiter.  

\cite{Zha20} suggested that shredding by low mass main sequence stars or white dwarfs could account for the interloper population. The Roche lobe of a star is only a few times larger than its radius, meaning that any star-shredded comet is likely to be strongly heated and at least partially devolatilized \citep{Ray18}, while comets shredded by giant planets can escape this fate.    A problem for this mechanism is already illustrated by the case of comet P/Shoemaker-Levy 9; most fragments will not escape.  \cite{Zha20} also proposed that shredding could match the elongated shape and devolatilized nature of `Oumuamua, before, however, estimates of the shape were revised to favor a disk-like body instead of a prolate one, and before the detection of non-gravitational acceleration suggested unseen outgassing of embedded volatiles.  Still, given that stars and planets can scatter objects to  interstellar space, the process must contribute, at some level, to the production of small interlopers from larger precursors.  Not  enough information about  other star and disk systems exists to meaningfully calculate the rate of production of objects by this process.

\textbf{Alien Technology:}
The prospect that 1I/`Oumuamua (but not 2I/Borisov) might represent evidence for alien technology has been so widely discussed that it deserves a mention here.  Two properties of `Oumuamua (its extreme shape and its non-gravitational acceleration in the absence of measurable outgassing)  are  unlike those of most solar system bodies.  But they are not unique to 1I/`Oumuamua.  For example, 2016 AK193 is a 0.1 km scale asteroid that reportedly shows a comparably large optical lightcurve (with a range of 3 magnitudes) indicative of an extreme shape \citep{Hei21}.  Likewise, other solar system bodies have recently been found to show non-gravitational acceleration in the absence of resolved evidence for outgassing  (\cite{Far23}, \cite{Sel23b}).  The non-gravitational acceleration of even the largest of these, the 300 m diameter asteroid (523599) 2003 RM, could be supplied by water sublimating at $\sim 2\times 10^{-3}$ kg s$^{-1}$, a minuscule rate that might easily escape detection.  Although `Oumuamua remains unique in having \textit{both} an extreme shape \textit{and} non-gravitational acceleration without coma, the high abundance of the interloper population raises other questions for the alien technology interpretation. With $\sim$10$^4$ similar objects inside Neptune's orbit at any one time, and an entering flux of $\sim$10$^3$ per year, it is reasonable to ask  why an alien civilization would send so many probes to the solar system and why, if their objective is to investigate Earth, would they miss by 0.16 au? \citep{Zuc22}.

\subsection{Interstellar Meteors and Interstellar Planets}
\textbf{Meteors:} To date, all well-established meteors have speeds relative to the Sun that are less than  $V_{e}$ = 42.1 km s$^{-1}$, the escape speed from the solar system measured at Earth's distance.  They consist of debris from asteroid collisions and material ejected from sublimating comets by gas drag.  However, the existence of sub-kilometer scale interlopers like 1I/`Oumuamua and 2I/Borisov implies that interstellar meteors caused by the impact of smaller counterparts must exist.   Contentious claims for the existence of faster (interstellar) meteors have been made for a very long time, based on both optical \citep{Opi50} and radar \citep{Bag07} reports.  The essential problem boils down to the difficulty in accurately measuring the speed, $U$, of a meteor during its limited flight time (typically a few $\times$0.1 s) in the air.  This difficulty is compounded by the uncertain effects of projection of the 3D path into the plane of the sky, and sometimes by physical effects including deceleration, ablation and fragmentation resulting from friction and ram pressure with the air. The excess velocity indicating a hyperbolic trajectory is $V_{\infty} = (U^2 - V_e^2)^{1/2}$ and, unless $U \gg V_e$, it is difficult to be confident that $V_{\infty}$ is significantly larger than the systematic errors of measurement \citep{Haj20}.

\cite{Sir22} report that fireball USG 20140108 had a velocity at infinity $V_{\infty}$ = 42.1$\pm$5.5 km s$^{-1}$ and that the responsible bolide penetrated to an altitude 18.7 km, an unusually low value indicating high material strength.  Unfortunately, the data behind this report are from a military satellite for which the details of measurement and their uncertainties are classified.  An independent analysis \citep{Bro23} (see also \cite{Vau22}) finds that these uncertainties are likely to be 10 to 15 km s$^{-1}$, in which case the excess velocity may not be significantly different from zero, and the trajectory may not be hyperbolic at all.  \cite{Bro23} find a plausible solution for a more typical low speed ($<$20 km s$^{-1}$) rocky bolide and this solution has the advantage of fitting the lightcurve of USG 20140108 without the need to assume extreme physical properties.  The diameter of USG 20140108 is $\sim$0.9 m, about 10$^3$ times the size (and 10$^9$ times the mass) of typical optical meteors.  It would be expected that numerous smaller examples of hyperbolic meteors should be found in optical and radar (c.f.~\cite{Bag07}) telescopic surveys whereas, in fact, there are no reliable cases \citep{Mus12}.  The point is that, while there is every reason to expect that interstellar meteors exist, there is as yet no unambiguous evidence that they do.  

In contrast, solid evidence exists that much smaller interstellar solids enter the planetary region, in the form of (mostly) micron sized and smaller dust particles carried in the neutral Helium stream (see \cite{Kru15} for a useful overview).  The coupling length of a 1 micron sized particle in the interstellar gas is $\sim$ 100 pc, so such particles follow the mean flow of gas on scales much larger than the average distance between stars.  The particles are too small to create optical meteors but are reliably identified by their high speed impact into  spacecraft equipped with time-of-flight detectors. They follow a roughly radius$^{-3}$ size distribution except that the smallest particles, radii $\le$0.1 $\mu$m, are repulsed by the action of solar radiation pressure, causing a downturn in the size distribution.   

\textbf{Impacts:} The interval between `Oumuamua-like impacts into the Earth, assuming a number density $\sim$0.1 au$^{-3}$, is $\sim$100 Myr to 200 Myr, so that over the 4.6 Gyr age of the solar system there may have been only $\sim$25 to 50 such impacts \citep{Jew20b}.   (For comparison, the impact interval for mass extinction-causing 10 km asteroids - each 10$^2$ times the size and 10$^6$ times the mass of 'Oumuamua - is $\sim$100 Myr). The  high relative speeds characteristic of hyperbolic orbits make it unlikely that any large fragments would reach the surface, reducing the prospect for finding large interstellar meteorites.  Indeed, most interstellar impactors would detonate at altitude as airbursts, leaving only dust to fall.  Conceivably, interstellar impactors could deliver biological material in this dust. For example, some terrestrial bacteria are known to survive long-term exposure to the space environment, and ``lithopanspermia'' is a viable mechanism for the transfer of living cells between planets \citep{Bel12}.  Experiments also show that terrestrial bacteria and lichen can survive strong shocks (several$\times10^{10}$ N m$^{-2}$) resulting from planetary impact \citep{Hor08}.  Given sufficient shielding from cosmic rays within an interstellar boulder, it seems possible that life could spread between the stars (\cite{Ada05}; \cite{Ada22}), or even galaxy-wide, by this mechanism.  This is a possibility for which, however, no evidence currently exists.

\begin{figure}
\centering
\includegraphics[scale=.4]{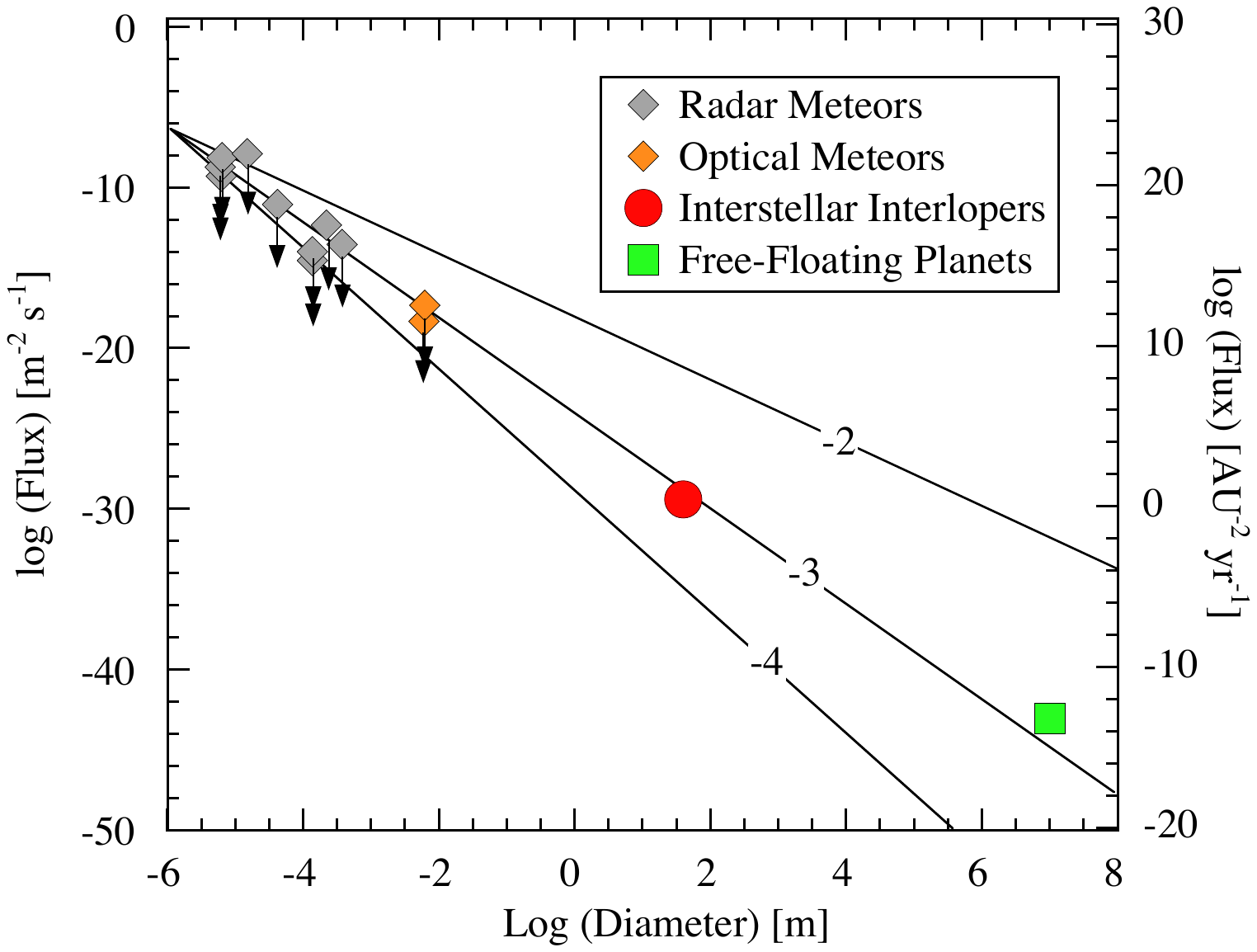}
\caption{Flux of objects, $F$, through the solar system as a function of their diameter, $D$.  Points plotted are for (upper limits to) radar and optical meteors, and estimates for interstellar interlopers (scaled from 'Oumuamua) and for free-floating planets.  For the latter, we used the number of FFPs per star from \cite{Sum23}, approximated the galaxy as a disk with radius 10 kpc and thickness 10 kpc, respectively, and assumed a velocity dispersion of 30 km s$^{-1}$.  For the other classes we used the data compiled by \cite{Jew23}. Three lines indicate $F \propto D^{\gamma}$, with $\gamma$ = -2, -3 and -4, as marked.  The interloper and FFP points are uncertain to order of magnitude but, given the large vertical scale of this plot, remain close to the  $\gamma$ = -3 line. Interstellar submicron particles have been measured but are partially shielded by radiation pressure and are omitted from the plot.  }
\label{iso}       
\end{figure}

\textbf{Planets:}  The gravitational processes responsible for the ejection of macroscopic bodies from planetary systems are independent of size over a wide range \citep{Ras96}. Ejected interstellar bodies much larger than the known examples, perhaps as large as planets, should exist, as might planets trapped in the Oort cloud (\cite{Ray23}), albeit much more rarely.  Indeed, interstellar (so-called ``free-floating'') planets (FFPs) have long been known principally from their gravitational microlensing effects on the brightness of more distant stars (e.g.~\cite{Mro18}) but also from direct imaging (e.g.~\cite{Pea23}).   Early estimates of the mass distribution function from gravitational lensing detections  suggest 21$_{-13}^{+23}$ FFPs per star, with a mass per star, 80$_{-47}^{+73}$ M$_{\oplus}$ \citep{Sum23}.  While very uncertain, this mass is comparable to or greater than the estimated mass/solar mass ejected to the interstellar medium during the Oort cloud formation phase of the solar system ($\sim$10 M$_{\oplus}$) and far larger than the $\sim$0.1 M$_{\oplus}$ per star in `Oumuamua-like bodies.  It is intriguing to think that interlopers and free-floaters differing in scale and mass by 5 and 15 orders of magnitude, respectively, might be connected by a single process (namely, gravitational scattering by planets). Figure \ref{iso})shows that interlopers and FFPs can be connected by a power law with index $\gamma \sim$ -3, and this index is even  consistent with observational upper limits to the population of interstellar meteors in the micron to millimeter range. On the other hand, the recent observation that 9\% of 540 free-floating Jupiter mass planets in the Trapezium cluster are resolved binaries brings into question our understanding of the ejection process, because violent ejection seems likely to split binaries \citep{Pea23}.  Future survey measurements will show whether or not the interlopers and the free floaters both belong to an as-yet dimly perceived but physically related  mass spectrum.

\section{The Future}
Improvement in the understanding of the interstellar interlopers largely depends on the identification of new examples.  The 8.4 m diameter telescope of the VRO (Vera Rubin Observatory, formerly known as the Large Synoptic Survey Telescope),  should play a major role, by repeatedly surveying the sky at optical wavelengths to greater depth than has so far been possible with smaller telescopes.  Estimates of the likely discovery rates are uncertain (both because the population is poorly known and because the operational parameters of VRO are still not empirically known), but plausible values based on scaling from `Oumuamua are $\sim10^{1\pm1}$ per year  (\cite{Hoo22}; \cite{Mar23}).  New detections are also expected from NASA's Near Earth Object Surveyor mission, which will use thermal infrared radiation in the 4 to 5 $\mu$m and 6 to 10 $\mu$m wavelength bands to study small bodies in the inner solar system \citep{Mai23}.  Data from VRO and NeoSurveyor will be complimentary, in the sense that optical data only provide a measure of the product of the albedo with the scattering cross-section, while the infrared flux effectively determines the cross-section (and hence the size) alone. 

Each new object will be subjected to the same intense observational onslaught as the first two, using telescopes around the world.   The central question posed by observations of 1I/`Oumuamua and 2I/Borisov is ``why are they so different?'', with 1I/`Oumuamua appearing unexpectedly aspherical and asteroidal and 2I/Borisov showing the morphological and spectroscopic properties more typical of a solar system comet.  Do these differences represent different evolutionary states of the same type of body (e.g.~is 1I/`Oumuamua an evolved version of 2I/Borisov from which the volatiles have been largely lost?), or are the differences intrinsic and related to different source regions or formation mechanisms?    New detections will show which type is the more common and may help to elucidate their possible inter-relationships.  New observations may also set `Oumuamua in better context, with implications for the origin of its extreme shape and non-gravitational acceleration.  Crucially, the new survey statistics will allow for the first time a meaningful assessment of the interloper magnitude and size distributions, so permitting better estimates of the size-dependent population and the total interloper mass.  The orbital element statistics (e.g.~are their orbits isotropically distributed or concentrated towards the solar apex, or towards nearby star formation regions?) will additionally set constraints on the sources of these objects.  

There has been considerable discussion about prospects for sending a spacecraft to an interstellar interloper, with the scientific motivation being to sample material accreted in another planetary system.  Such a mission will be extremely challenging, both because interlopers will invariably be discovered only shortly before their closest approach and because the encounter speeds are necessarily very high.  ESA's Comet Interceptor mission \citep{Sno19}, with launch expected in 2029, will lurk at the L2 Lagrangian point waiting for a suitable target.  Having very weak ion drive propulsion, Comet Interceptor is  unlikely to be within reach of an interstellar interloper in the six year duration of its mission \citep{Hoo22}, but it serves as a model of the type of preparation that will be needed to make a rendezvous eventually possible.

\section{Bibliography}

\section{Cross-References}

\begin{itemize}
\item{The Diverse Population of Small Bodies of the Solar System}
\item{Brown Dwarfs and Free-Floating Planets in Young Stellar Clusters}
\item{Dynamical Evolution of Planetary Systems}
\end{itemize}



\end{document}